# A Simple Prediction Model for the Development Trend of 2019-nCov Epidemics Based on Medical Observations

**Ye Liang[1,2], Dan Xu[4], Shang Fu[6], Kewa Gao[3,5], Jingjing Huan[2], Linyong Xu[2], Jia-da Li[2*]**

1 Center of Stomatology, Xiangya Hospital, Central South University, Changsha, China
2 School of Life Sciences, Central South University, Changsha, Hunan Province, China
3 The Bioengineering Laboratory of University of California, Davis Medical Center, Sacramento, CA, United State
4 GF Securities Asset Management (Guangdong) Co., Ltd, Guangzhou, China
5 The Third Hospital of Central South University, Hunan, Changsha, China
6 Department of material science and engineering. Central South University

*Corresponding author
Jia-Da Li
E-mail address: Lijiada@sklmg.edu.cn

First author
Ye Liang
E-mail address: Liangye@csu.edu.cn

## Abstract

In order to predict the development trend of the 2019 coronavirus (2019-nCov), we established an prediction model to predict the number of diagnoses case in China except Hubei Province. From January 25 to January 29, 2020, we optimized 6 prediction models, 5 of them based on the number of medical observations to predicts the peak time of confirmed diagnosis will appear on the period of morning of January 29 from 24:00 to February 2 before 5 o'clock 24:00. Then we tracked the data from 24 o'clock on January 29 to 24 o'clock on January 31, and found that the predicted value of the data on the 3rd has a small deviation from the actual value, and the actual value has always remained within the range predicted by the comprehensive prediction model 6. Therefore we discloses this finding and will continue to track whether this pattern can be maintained for longer. We believe that the changes medical observation case number may help to judge the trend of the epidemic situation in advance.

# Background:

The 2019 coronavirus (2019-nCov) is currently raging in China and has a tendency to spread to the world. The number of infections has skyrocketed since the first announcement on December 31, 2019, and has spread to all provinces in China within 30 days. However, the changing trend of transmission and the time when the climax comes still not clear yet. A Correct prediction of the development trend of the epidemic is of great significance to reduce the total number of infections and prevention of new infections and control disease spread. In this study, we established a prediction model by analyzing the numerical regularity of the data and the logical association of the indicators according to publicly reported data to predict disease progression. It will contribute to the development of future strategies for disease prevention and control.

# Overall hypothesis

**Assumption one:**

The human trajectory is in a random state,the virus does not mutate, and the probability of different patients infecting other people in the same period of time is close.

**Assumption two:**

Hubei Province is the place where proband cases occurred. In the early days of disaster relief,resources were insufficient. Later, The government adopt a strict strategy that limit citizen's mobility in Hubei province,including Wuhan city. This study intends to study the data of Hubei Province, study the data of other provinces in China except Hubei Province, and infer the disaster trend outside Hubei Province.

**Assumption three:**

The infection cases outside Hubei province are homologous, the diagnosis standards are uniform across the provinces, and the confirmed data are reliable and the data are regular.

**Assumption four:**

Most of the confirmedcases experienced medical observation at first.

# Result

The results of the final prediction model established on January 29 based on the data on January 28 at 24 o'clock are shown in Figure 1:

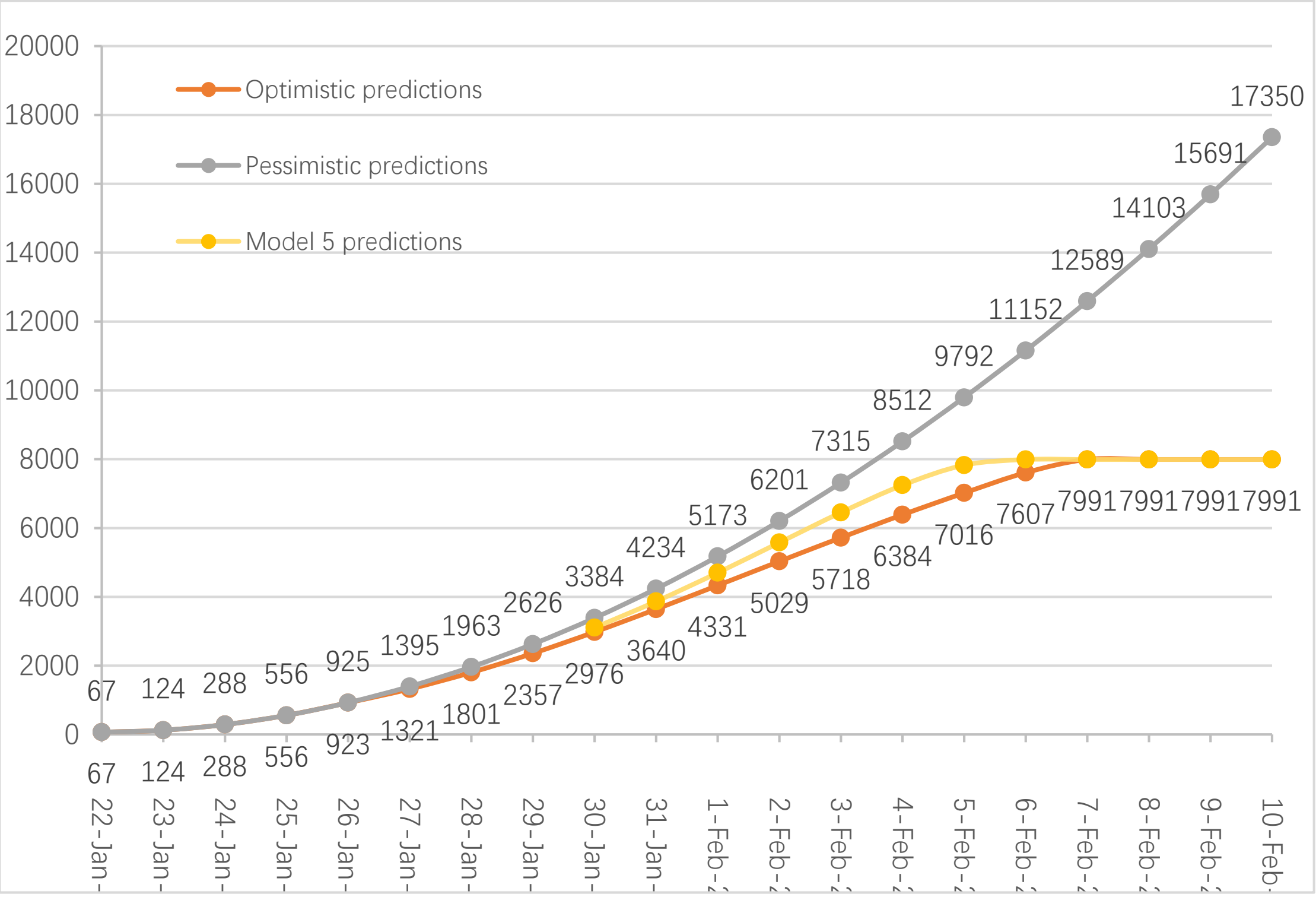

Figure 1. Final prediction model results

In order to verify the validity of the model, we observed the 24 o'clock data for three days from January 29 to 31, 2020, all of which were within the forecast range, and tended to be optimistic. The actual measurement situation is shown in Figure 2:

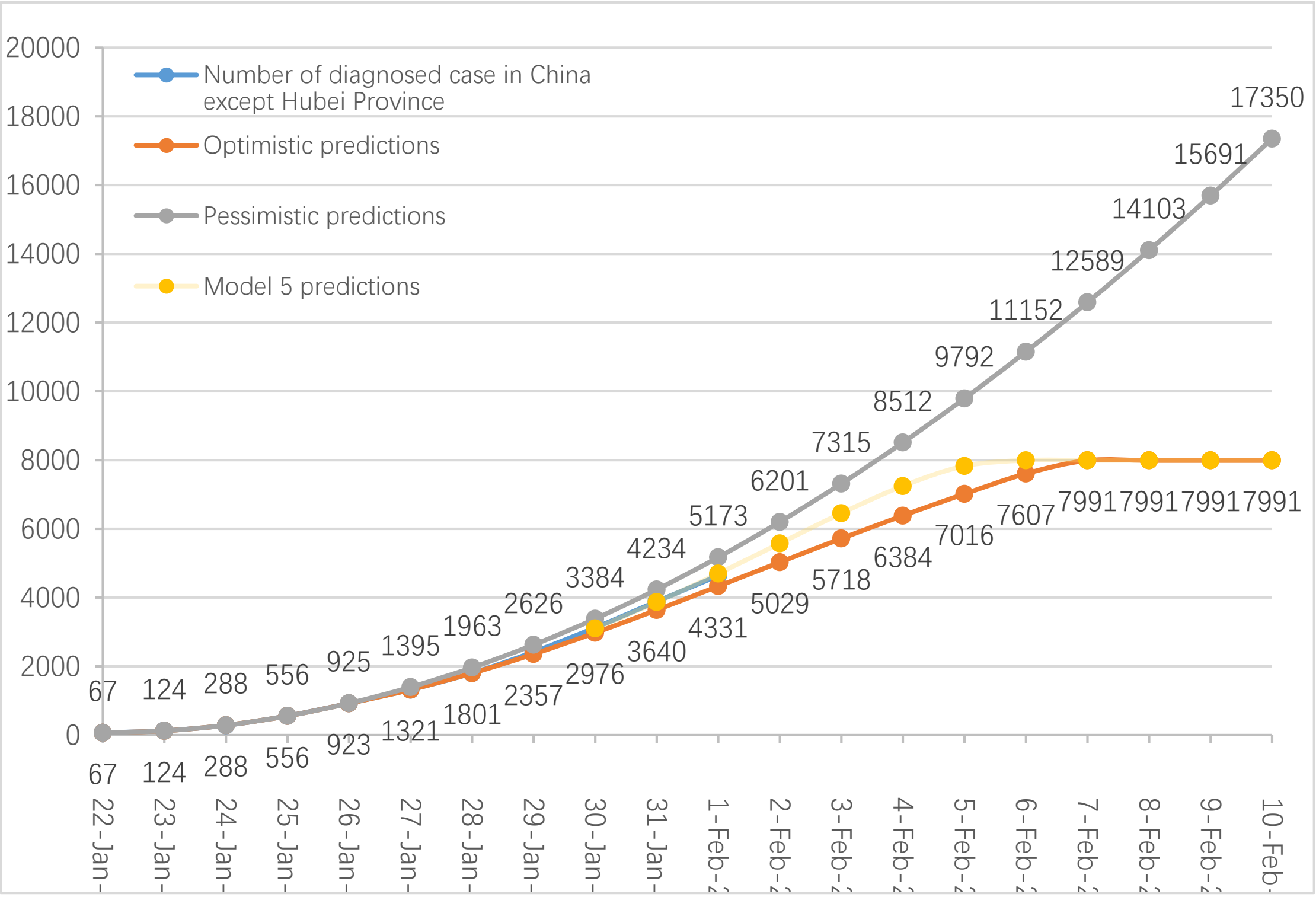

Figure 2. Comparison of final prediction model results with actual measurements

# Method

## Symbol Description

$C_i$ : Case number of patient in China at day i
$H_i$ : Case number of patient in Hubei at day i
$E_i$ : Case number of patient in China except of Hubei at day i
$T_i$ : Case number of close contact with patient
$R_i$ : Case number of released from medical observation
$O_i$ : Case number of medical observation
$\varphi$ : Integration initial parameters
$\omega$ : Integral increment parameter
$\delta$ : Initial value related to newly diagnose
$\mu$ : Variation Coefficient of related to newly diagnose
$\varepsilon$ : Ratio of medical observation and newly diagnose
$\sigma$ : Variation Coefficient of ratio of medical observation and newly diagnose
$\tau$ : Amplification coefficient of case leased from medical observation case
$\beta$ : Increasing rate of medical observation case
$\gamma$ : Increasing rate of case leased from medical observation case

Data source and sampling instructions
The national data comes from the Chinese center for disease control and prevention [1], in this study recorded as $C_i, E_i, T_i, R_i, O_i$. The data of Hubei comes from the Health Commission of Hubei Provice [2], in this study recorded as $H_i$. Both data sources released the outbreak data at 24:00 on the previous day that morning. Scattered daily data of other channels are not counted. Therefore, with the exception of Hubei Province, the total number of daily diagnoses in China is: $E_i = C_i - H_i$ .

## Model 1: prediction based on fitting of confirmed data

The model was established on January 25, 2020. The model fits the data before January 25 by polynomial method, and infers the subsequent data. The fitting equation is:

$$y = -\frac{1}{3}x^3 + 55.071x^2 - 105.6x + 116.8$$

where the data at 24 o'clock on January 25 is predicted to be x = 5. The prediction curve is shown in Fig.3

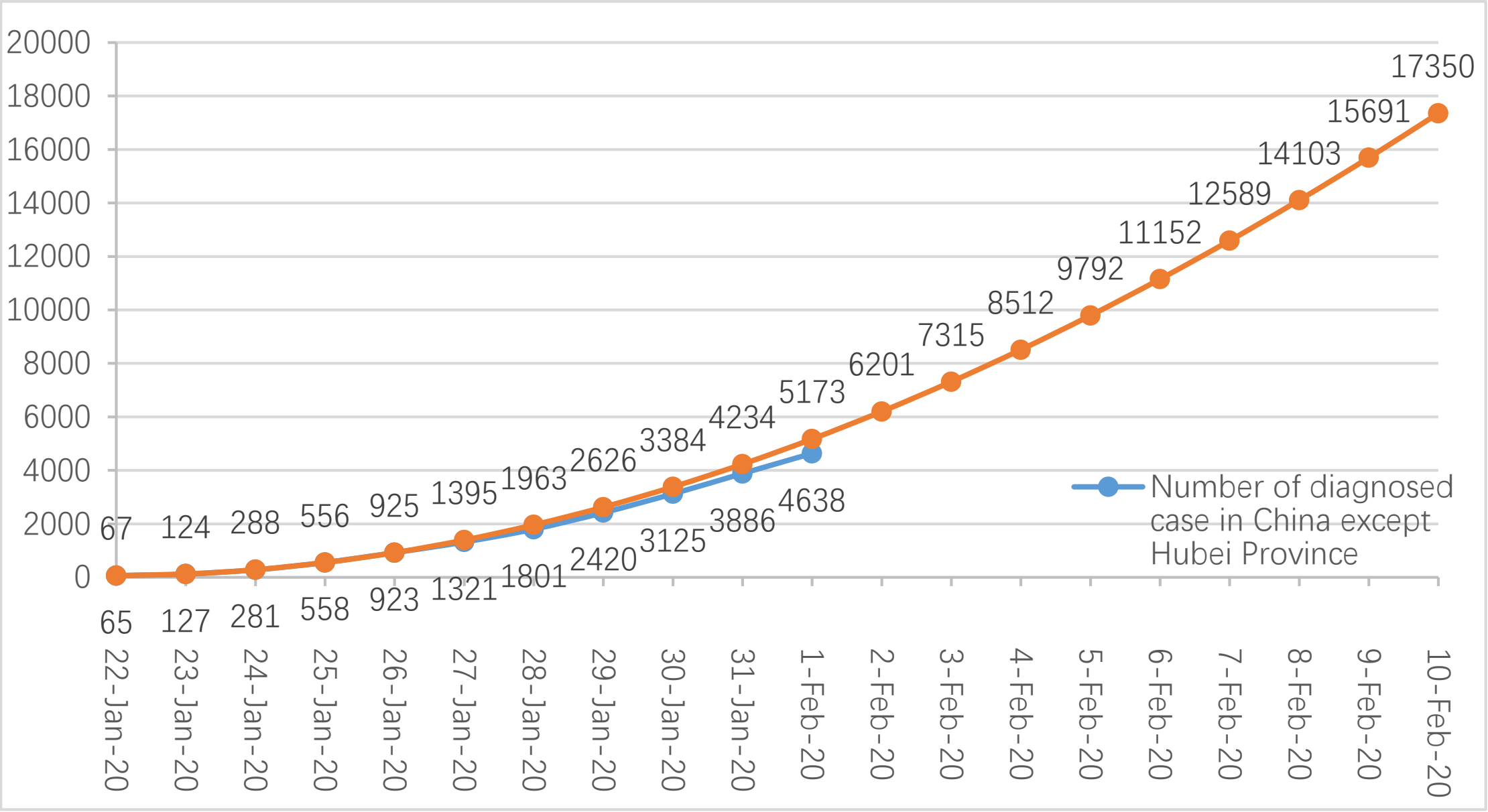

Figure 3. Comparison of prediction model 1 with actual measurements

## Model 2: Superimposed prediction based on confirmed data and actual deviation

After the first model was established, the three data at 24:00 on January 25-27 were analyzed, and it was found that the actual data gradually deviated from the model. Therefore, the newly obtained data is used for the secondary correction. When the correction method is used, the deviation is fitted twice. The fitting equation is:

$$\Delta y = -8x^2 - 48x + 54$$

where the data at 24 o'clock on January 25th is predicted to be x = 1. It will be superimposed with model one. To get model two. The predicted image is shown in Figure 4.

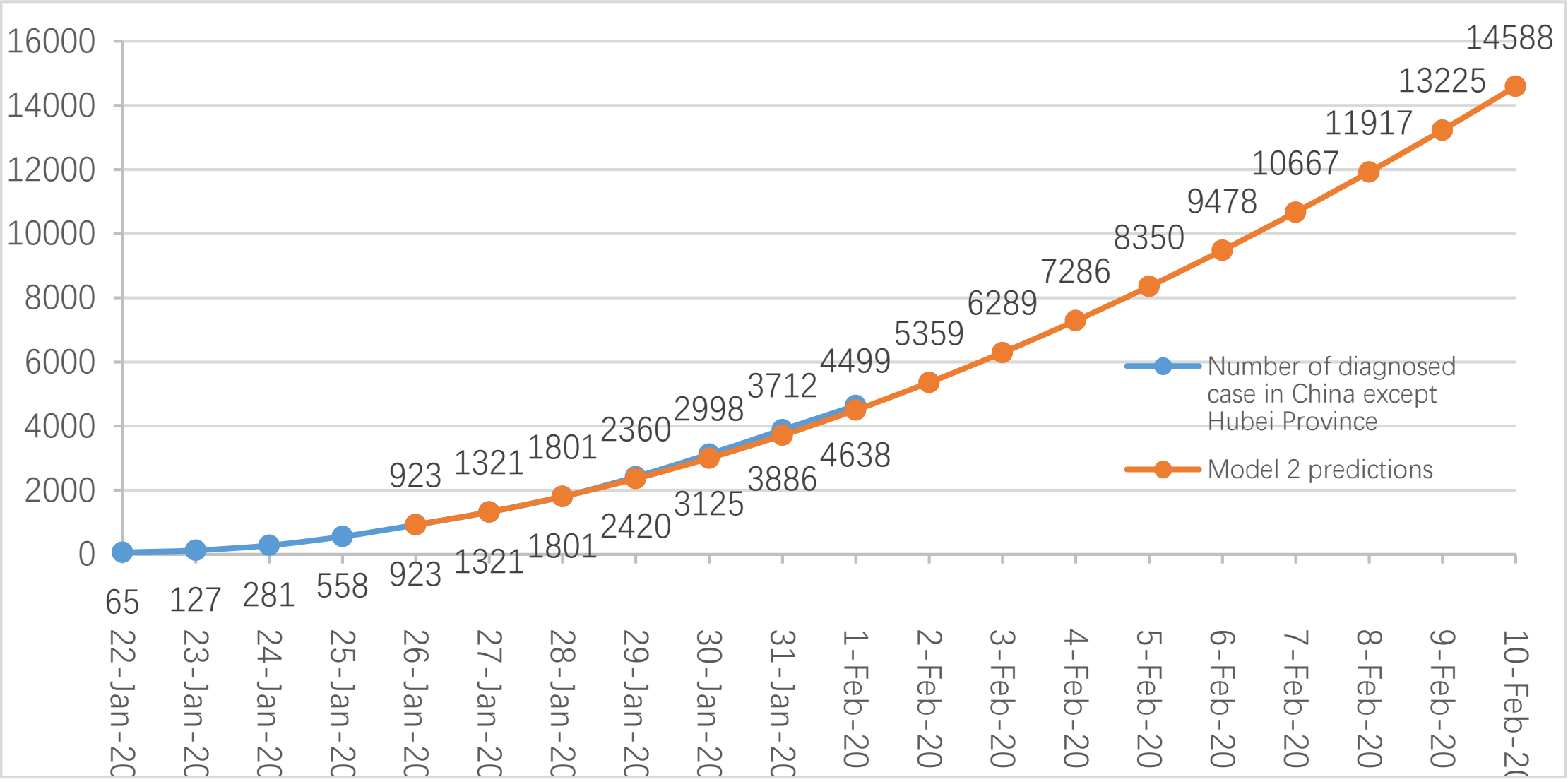

Figure 4. Comparison of prediction model 2 with actual measurements

## Model 3: integral prediction model

The model was established on January 28. The model assumes that there is a correlation between the daily increase in the number of confirmed diagnoses and the number of confirmed diagnoses in the previous day. Based on this assumption, the model expression is derived:

$$E_n = E_0 + n\varphi + \omega\sum_{i=1}^{n} i$$

The model simulation results after calculating the parameters using the data before January 27th at 24:00 are shown in Figure 5:

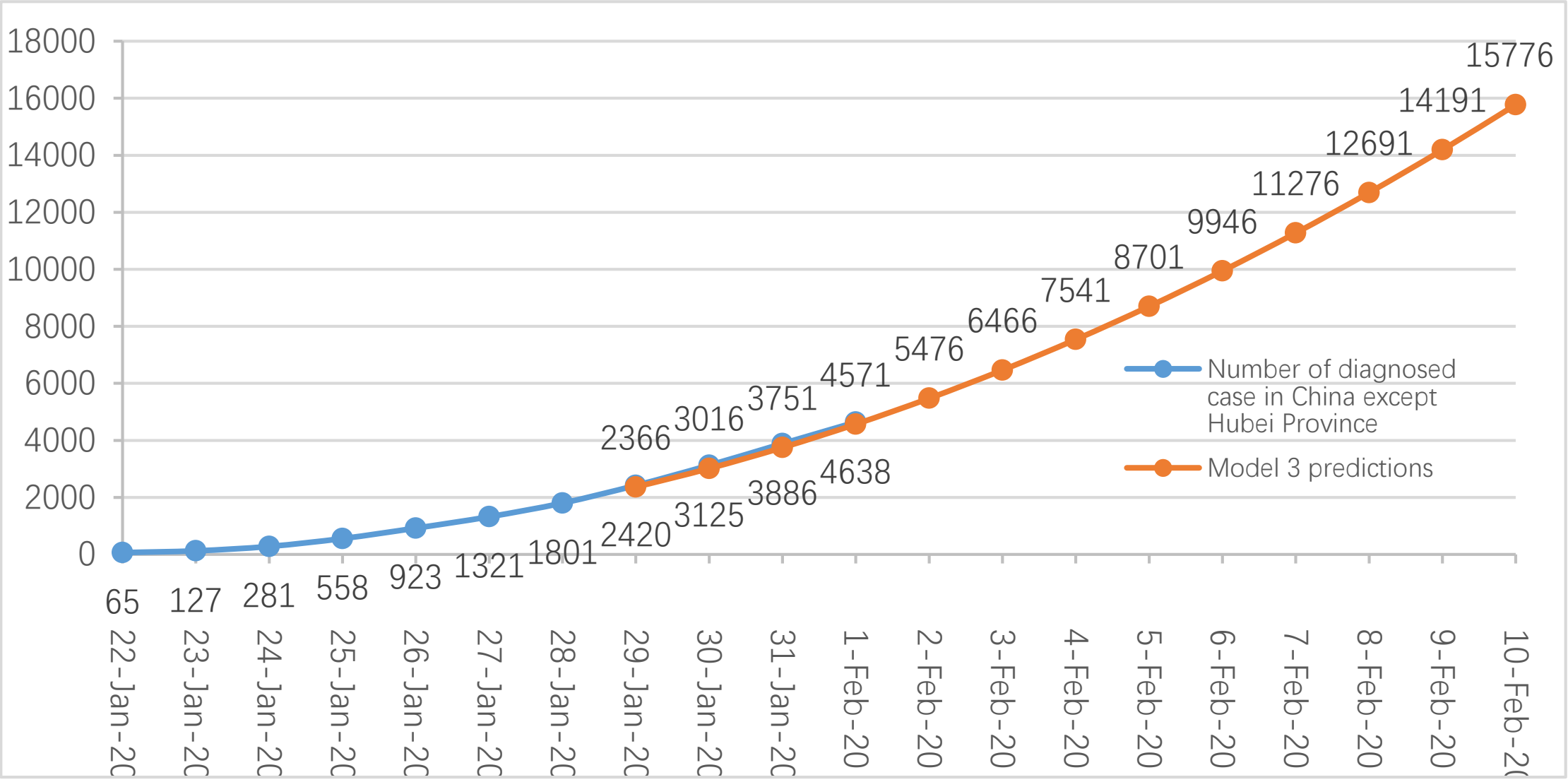

Figure 5. Comparison of prediction model 3 with actual measurements

## Model 4: Correlation prediction between existing diagnoses and new people

The model was established on January 28. The model assumes that there is a correlation between the daily increase in the number of confirmed diagnoses and the total number of diagnoses. This correlation can be used to establish a mathematical model. From this, the model expression is derived:

$$E_n = E_0 \prod_{i=1}^{n} (1 + \mu^i P_0)$$

Figure 6 shows the model simulation results after using the data before January 27th to calculate the parameters:

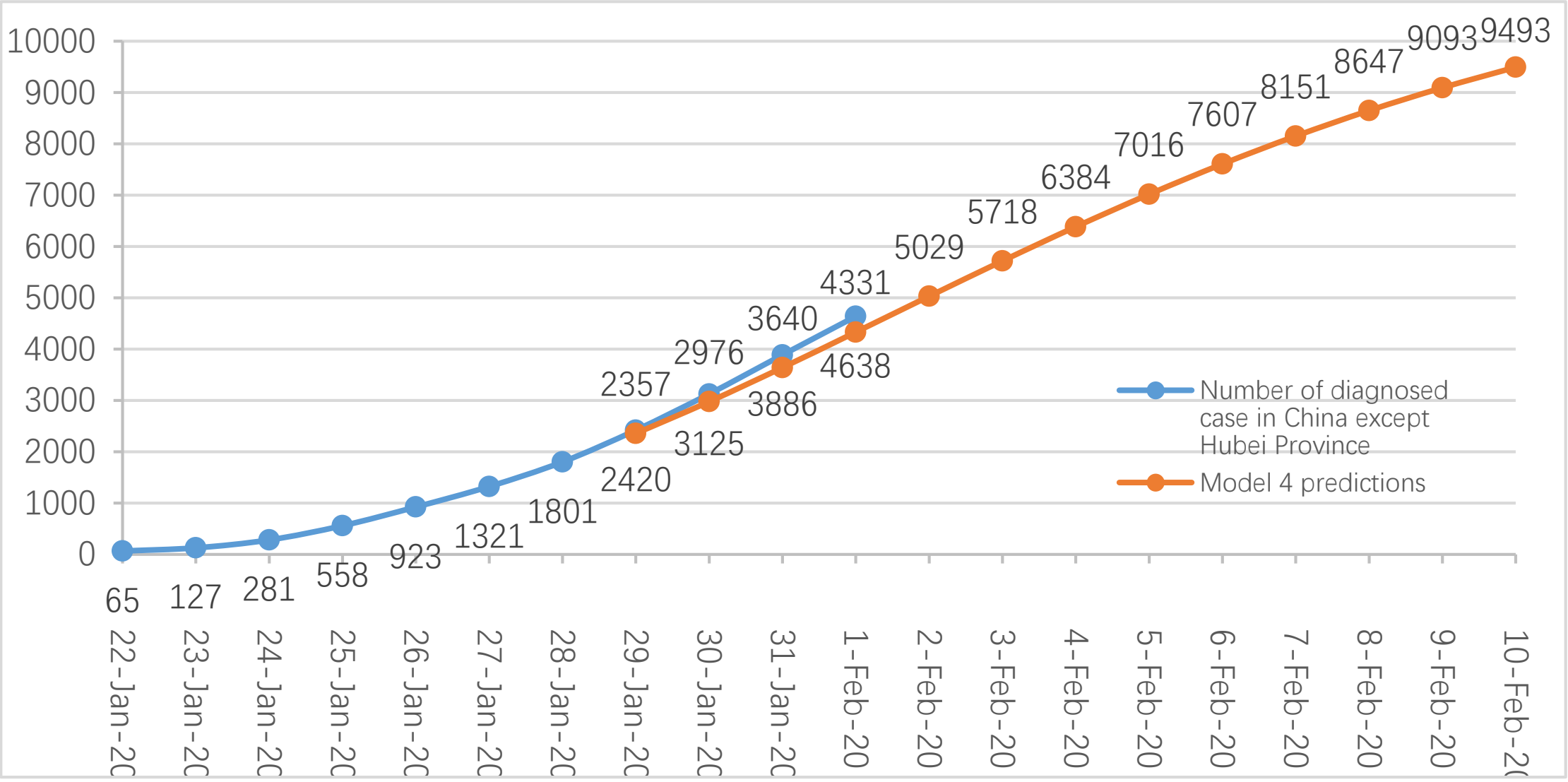

Figure 6. Comparison of prediction model 4 with actual measurements

## Model 5： Prediction model based on the case number of medical observations

The model was established on January 28. The following assumptions been added in this model.

- There are relationship between the number of new diagnose case and the number of people undergoing medical observation the previous day. The daily correlation can be described by $\varepsilon_i$, and the variation of $\varepsilon_i$ can be described by $\sigma$.
- There is relationship between the number of people undergoing medical observation each day and the number of medical observations on the previous day and the number of people who released from medical observations. The correlation can be described by $\gamma, \tau$;
- There is relationship between the number of daily medical observations released and the number of medical observations released the day before;
- The number of people being observed will not be less than 0.

It is deduced that the number of medical observations on the n$^{th}$ day conforms to the following model:

$$O_n = r^n O_0 - \tau R_0 \sum_{i=1}^{n} r^{n-i} \beta^i$$

The proportion of medical observations that are diagnosed daily can be calculated by model

as follow:

$$\varepsilon_n = \sigma^n \varepsilon_0$$

The patient increase on day i corresponds to:

$$\Delta_\mathrm{i} = \max(\varepsilon_\mathrm{i} O_{i-1}\ , 0)$$

Therefore, the patient on day i + 1 can be recursively calculated using the following formula:

$$\mathrm{E}_{\mathrm{i}+1} = E_i + \Delta_\mathrm{i}$$

The model simulation results after calculating the parameters using the data before 24:00 on January 27 are shown in Figure 7:

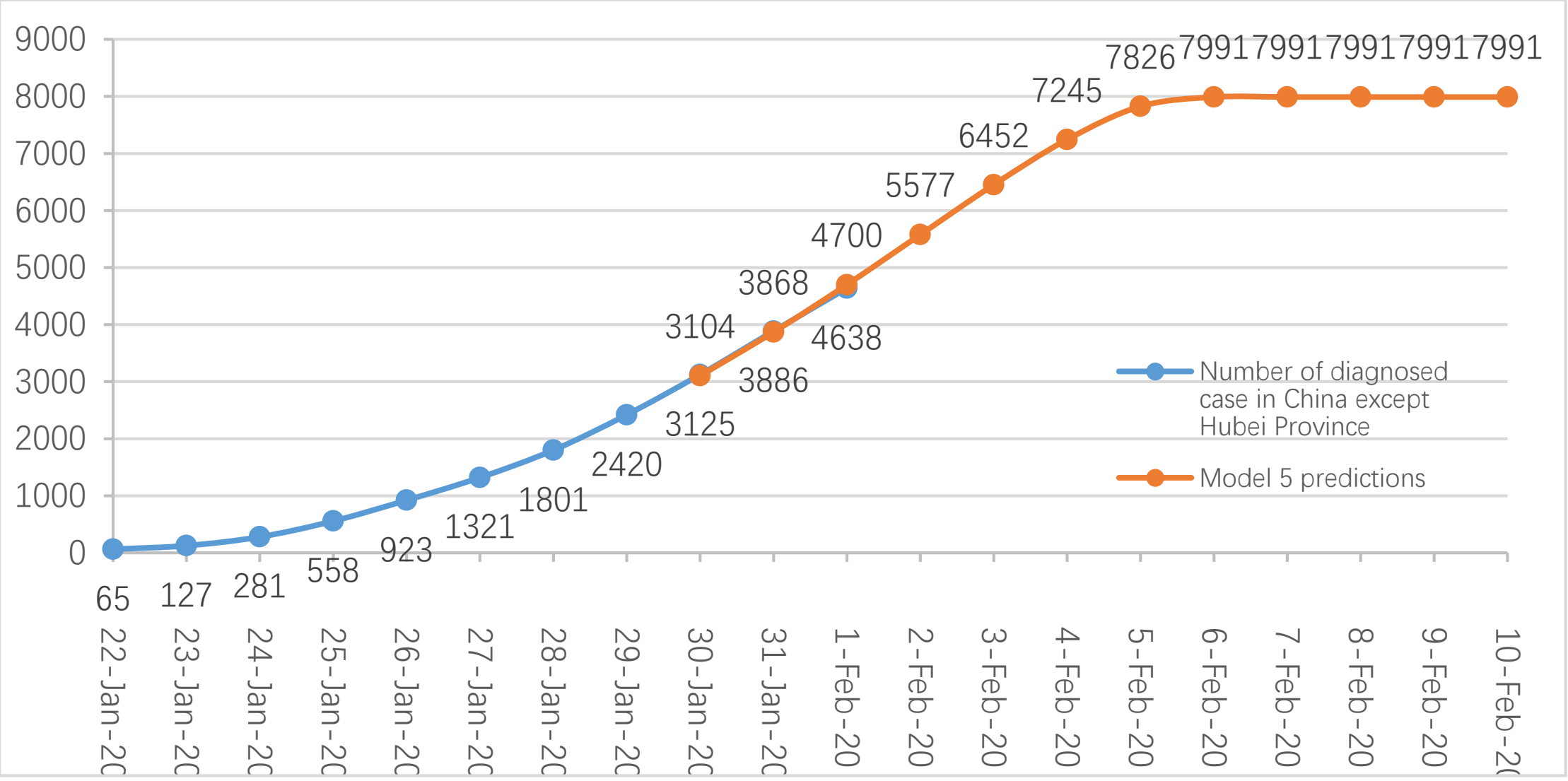

Figure7. Comparison of prediction model 5 with actual measurements

# Model 6 comprehensive prediction model

The prediction results of the aforementioned five models are similar. In order to obtain the overall situation consideration, the minimum value of the above model is used to obtain the optimistic forecast data, and the maximum value is used to obtain the pessimistic forecast data to comprehensively judge the trend of the epidemic situation. The predictions of this model are shown in the results section.

# Conclusion

This article shows some basic mathematical models established by numerical rules. Among them, the model 5 based on the number of medical observations predicts that the peak number of confirmed patients will appear between 24:00 on February 2 and 24:00 on February 5. This may help to determine the trend of the epidemic.

# Acknowledgements:

At the moment of the catastrophe, first of all, I would like to thank all the doctors and fighters in the clinical front-line. Secondly, I would like to thank all readers. In order to find out the development of the epidemic as soon as possible and encourage more people to join the scientific research and judge the epidemic, the text and graphics of this article have not been adjusted to the perfect state. Welcome to contact and correct. Special thanks to Professor Changyun Fang, Guoqing Huang, Juan Du, Yupeng liu, Ang Deng, Feiyue Zeng, and Shaohua Liu for their continuous attention and supervision of the prediction process, and to Prof. Zhoushun Zheng and Prof. Canhu Jiang a for their support in all aspects.

## Data Sources